\documentclass[%
 reprint,
superscriptaddress,
 amsmath,amssymb,
 aps,
]{revtex4-2}

\usepackage{graphicx}
\usepackage{dcolumn}
\usepackage{bm}

\usepackage{xcolor}
\usepackage{soul}
\usepackage{upgreek}

\renewcommand{\textcolor}[2]{#2}

\begin{document}

\preprint{APS/123-QED}

\title{Characterization of photon pairs generated by a silicon ring resonator under electrical self-pumping}

\author{Francesco Garrisi}
\email{francesco.garrisi01@universitadipavia.it}
 \affiliation{Dipartimento di Fisica, Universit\`a degli Studi di Pavia, via A. Bassi 6, 27100 Pavia, Italy}
\author{Federico Andrea Sabattoli}%
\affiliation{Dipartimento di Fisica, Universit\`a degli Studi di Pavia, via A. Bassi 6, 27100 Pavia, Italy}

\author{Nicola Bergamasco}
 \affiliation{Dipartimento di Fisica, Universit\`a degli Studi di Pavia, via A. Bassi 6, 27100 Pavia, Italy}
 
 \author{Micol Previde Massara}
 \affiliation{EPFL, STI IMT GR-QUACK, Station 11, CH-1015, Lausanne, Switzerland}

 \author{Federico Pirzio}
 \affiliation{Dipartimento di Ingegneria Industriale e dell'Informazione, Universit\`a degli Studi di Pavia, via A. Bassi 6, 27100 Pavia, Italy}

 \author{Francesco Morichetti}
 \affiliation{Dipartimento di Elettronica, Informazione e Bioingegneria, Politecnico di Milano, 20133 Milano, Italy}
 
 \author{Andrea Melloni}
 \affiliation{Dipartimento di Elettronica, Informazione e Bioingegneria, Politecnico di Milano, 20133 Milano, Italy}

 \author{Marco Liscidini}
 \affiliation{Dipartimento di Fisica, Universit\`a degli Studi di Pavia, via A. Bassi 6, 27100 Pavia, Italy}
 
 \author{Matteo Galli}
 \affiliation{Dipartimento di Fisica, Universit\`a degli Studi di Pavia, via A. Bassi 6, 27100 Pavia, Italy}
  
 \author{Daniele Bajoni}
 \affiliation{Dipartimento di Ingegneria Industriale e dell'Informazione, Universit\`a degli Studi di Pavia, via A. Bassi 6, 27100 Pavia, Italy}


\begin{abstract}
We report on the generation of nonclassical states of light in a silicon ring resonator in a self-pumping scheme. The ring is inserted in a lasing cavity, for which it acts as a filter, so that the lasing always occurs within a selected ring resonance, without active stabilization of the resonance frequency. We show the emission of coincident photon pairs and study their correlation properties through the reconstruction of the measurements via stimulated emission.
\end{abstract}

\maketitle


The development of efficient, reliable and cost effective sources of quantum states of light is of key importance for the widespread use of many quantum technologies \cite{Caspani_review}. Among these sources,
silicon ring resonators meet exactly these criteria, as they are microscopic, realized with mature fabrication processes, and demonstrate efficient optical nonlinearities at low optical powers \cite{Almeida2004,AlmeidaLipsonOL2004,Turner2008}. In these systems, spontaneous four-wave mixing (SFWM) can be used to realize efficient sources of photon pairs with emission rates in the MHz range \cite{Azzini2012OE,Clemmen2009}. Such pairs can be used to produce quantum states of light, in particular entangled photon pairs \cite{Grassani:Optica:2015,Wakabayashi:2015:ringtimebin,Harris:PRX:2014}. 

The advantage of using integrated sources is however reduced by the need of tunable lasers to be used as pump.
Resonant sources are generally needed for efficient photon pair emission, but the resonance frequencies are sensible to power and thermal fluctuations \cite{Carmon:2004,Peccianti:2012} so that the pump must often be actively controlled to accommodate for these shifts in frequency, particularly when dealing with high-Q resonators \cite{Jiang:Painter:2015:microdiskcoinc} or high pumping powers \cite{Azzini2012OL}.
These problems can be mitigated by using self-pumping schemes \cite{Pasquazi:2013,Reimer2014}, where the photonic chip containing the ring resonator used to generate photon pairs is inserted in a lasing cavity, and thus acts as a filter:
then, any shift of its optical spectrum will tune automatically the pump laser mode.
\textcolor{red}{In this work we self-pump a silicon ring resonator and we characterize the photon pairs generated via SFWM in terms of emission rates and of the degree of spectral correlation between the pairs.}
\begin{figure*}
    \centering
    \includegraphics[width=0.95\textwidth]{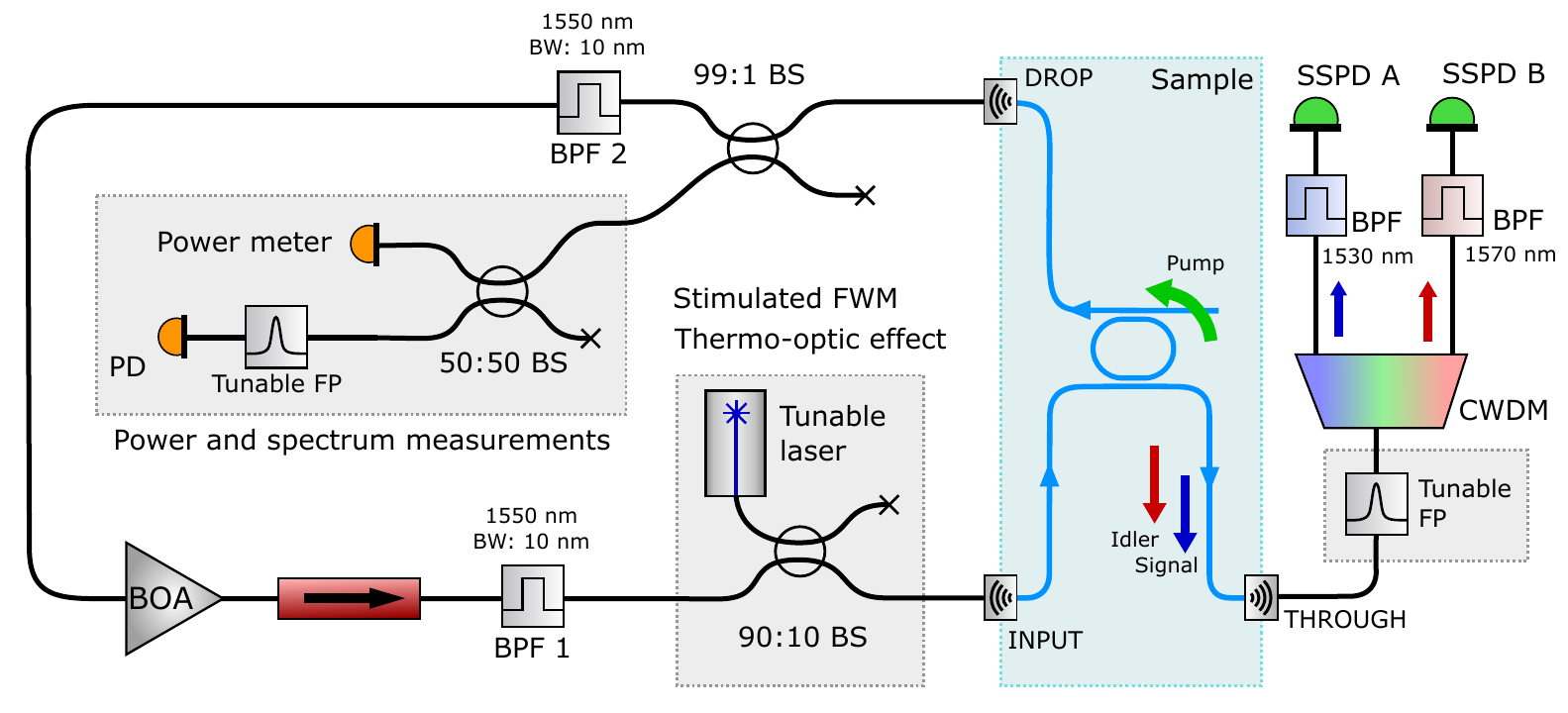}
    \caption{Experimental setup. An integrated microring resonator in the add-drop configuration is included in a loop cavity and acts as a filter, thus the booster optical amplifier provides gain for the radiation of one of the ring resonances to obtain lasing.
    Gray boxes contain additional parts of the set-up used for tests and characterizations.
    BS: Beam Splitter. PD: Photo Diode. SSPD: Superconducting Single-Photon Detector. BPF: Band-Pass Filter. BW: Bandwidth. CWDM: Coarse Wavelength Division Multiplexer.}
    \label{fig:setup}
\end{figure*}

The experimental setup is shown in Fig. \ref{fig:setup}.
A thermally stabilized sample fabricated in a commercial silicon on insulator (SOI) platform (Advanced Micro Foundry) contains a critically-coupled microring resonator in the add-drop configuration and is realized with silica buried waveguides whose cross-section is $500 \times 220$ nm$^2$.
The resonator has the shape of a racetrack, it is 66.8~$\upmu$m long and its free spectral range (FSR) is 8.6~nm, from which we infer that its group index is 4.18.
One resonance centered around $\lambda_p = 1547.6$ nm is chosen as the pump resonance, while the signal and idler resonances are two FSRs apart and are centered at $\lambda_s = 1530.4$ nm and $\lambda_i = 1564.9$. The quality factors of the three resonances are $22\,100$, $21\,900$ and $14\,900$, respectively.

Light is coupled into and out of the sample through grating couplers, with measured 9 dB total losses between the Input and Drop ports when on resonance with the microring.
The light exiting from the Drop port of the ring is externally routed using polarization maintaining components to the Input port, thus forming a closed loop cavity.
The loop includes a booster optical amplifier (BOA) to provide gain to the cavity field and to achieve lasing operation.
Along the cavity a 99:1 beam splitter (BS) is used to monitor the laser radiation.
Half of the tapped power is routed to a power meter (PM) to measure the cavity power level, while the other half is sent to a photodiode (PD) preceded by an electrically tunable fibre Fabry-P\'erot (FP) to assess the lasing spectrum.
The FSR of the FP is 240~pm while its full width at half maximum (FWHM) is smaller than 1~pm.
\textcolor{red}{Taking all losses into account, the power at the ring input waveguide ($P_\text{in}$) is estimated to be 565 times the power measured with the PM.}
The two band-pass filters (BPF) along the cavity are 10 nm wide and are centered on the pump resonance.
The filter at the output of the BOA (BPF 1) rejects the amplified spontaneous emission that would otherwise overcome the spontaneously generated photons in the microring, while the filter after the 99:1 BS  (BPF 2) prevents the amplification of light generated by SFWM in the ring.





\begin{figure}[h]
    \centering
    \includegraphics[width=0.43\textwidth]{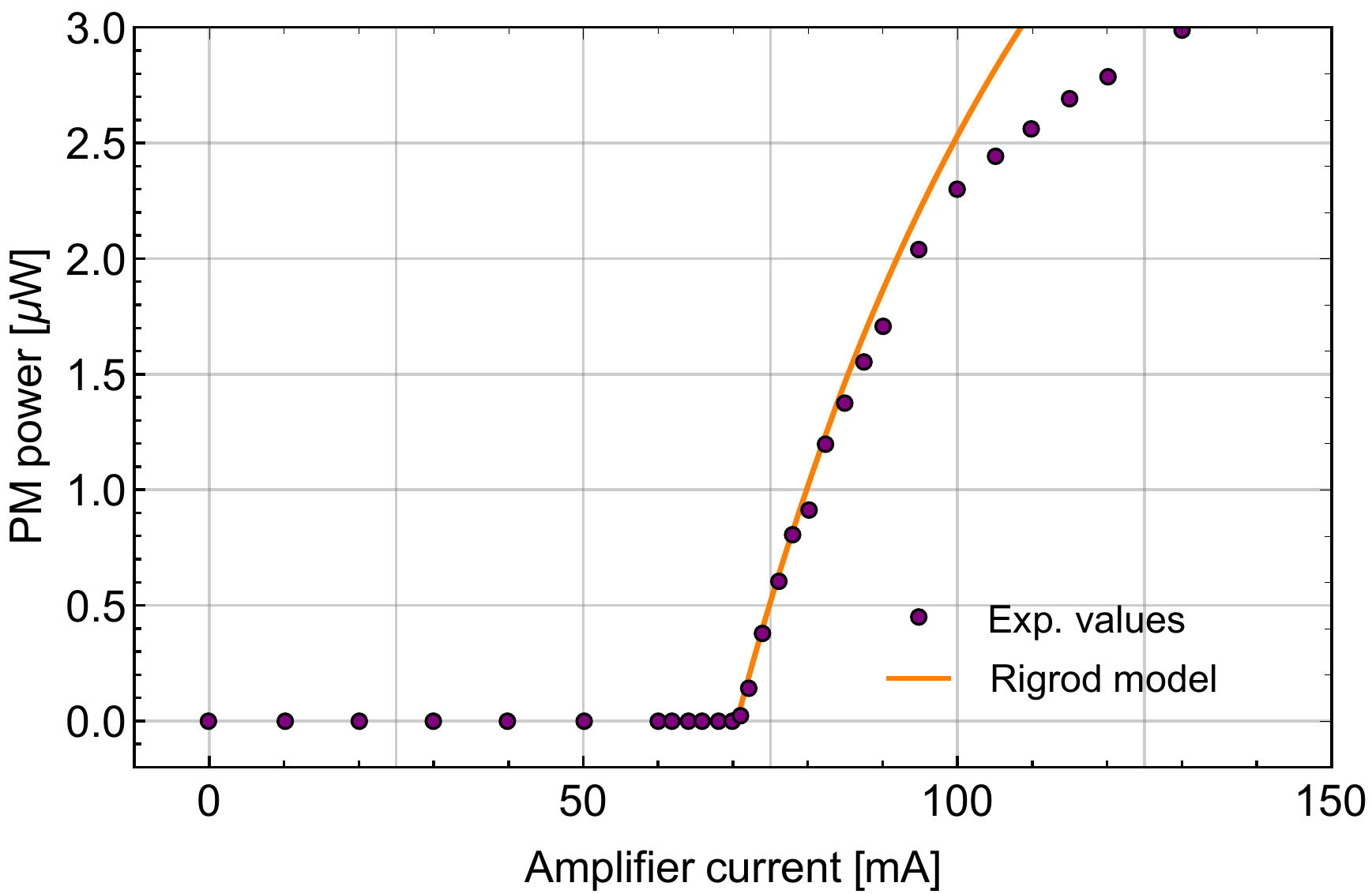}
    \caption{Lasing curve of the loop cavity. The curve is approximately linear above the threshold current ($I_\text{th} = 70.4$ mA) and follows well the theoretical model (orange line); the slope efficiency, $\eta = 100.1$ nW/mA, is obtained from a linear fit. Approximately above 100 mA the cavity behaviour deviates from the prediction due to nonlinear losses.}
    \label{fig:lasingcurve}
\end{figure}

In Figure \ref{fig:lasingcurve} we show the lasing curve of the cavity as a function of the current supplied to the amplifier.
We compare the experimental points with the Rigrod model \cite{rigrod1965saturation} (orange curve in Fig. \ref{fig:lasingcurve}) by taking into account the round-trip transmission of the cavity ($T_\text{tot}$), and the BOA small signal gain curve and saturation power (15 dBm).
The value of $T_\text{tot}$ is measured from the output of the amplifier to its input to be 8.95\% ($-10.5$~dB) and is mainly due to the coupling losses to the sample ($-9$~dB) and to the BPFs ($-0.5$~dB each); \textcolor{red}{the remaining -0.5 dB loss is due to optical connections along the loop.}
The model is in good agreement with the experimental points above threshold. At currents higher than 100~mA the theory deviates from the observed behaviour, likely due to excess losses introduced by the microresonator, as explained below.

\begin{figure}[h]
    \centering
    \includegraphics[width=0.45\textwidth]{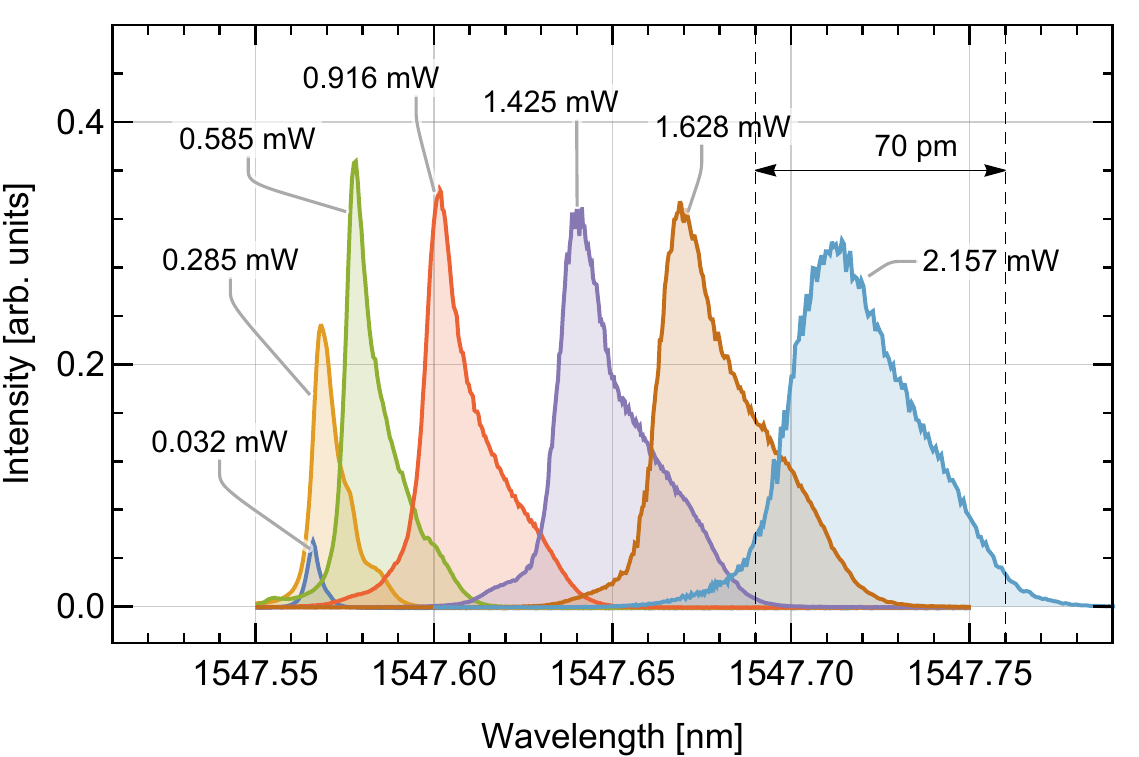}
    \caption{Spectrum of the cavity radiation at increasing ring input powers ($P_\text{in}$). The lasing radiation broadens, and it follows the ring resonance as it redshifts due to the thermo-optic effect. The Dashed vertical lines represent the FWHM of the cold ring pump resonance.}
    \label{fig:thermoshift}
\end{figure}

Thermo-optic effect and TPA play a key role in the behaviour of the cavity at high powers.
In Figure \ref{fig:thermoshift} we show lasing spectra for different cavity powers.
As expected, the lasing radiation follows the microring resonance as it redshifts due to the thermo-optic effect, maintaining a stable power level; lasing is preserved even when the resonance shifts by more than its FWHM.
At the same time, while at low powers the lasing line is narrow, at higher powers it broadens, becoming highly multimoded and comparable to the ring linewidth (\textasciitilde 70 pm).
\textcolor{red}{This effect is attributed to the interplay between the broadening of the ring resonance, due to increased looses induced by TPA, and gain saturation effects in the amplifier.}\\


The lasing radiation circulating in the fiber loop cavity is used to generate photon pairs in the the microring by SFWM.
To characterize the process, we measure the coincidence rate between signal and idler photons at the Through port of the microring (Fig.~\ref{fig:setup}).
In this configuration we have a pump suppression of about 15~dB, as the ring is close to the critical coupling condition.
A coarse wavelength division multiplexer (CWDM), with equally spaced 20 nm channels, is used to spatially separate signal and idler and to further suppress the pump.
Additional BPFs are necessary to completely reject the pump field before sending the pairs to high-efficiency (\textasciitilde 80\%) superconducting single-photon detectors (SSPD) for correlation measurements.

\begin{figure}
    \centering
    \includegraphics[width=0.48\textwidth]{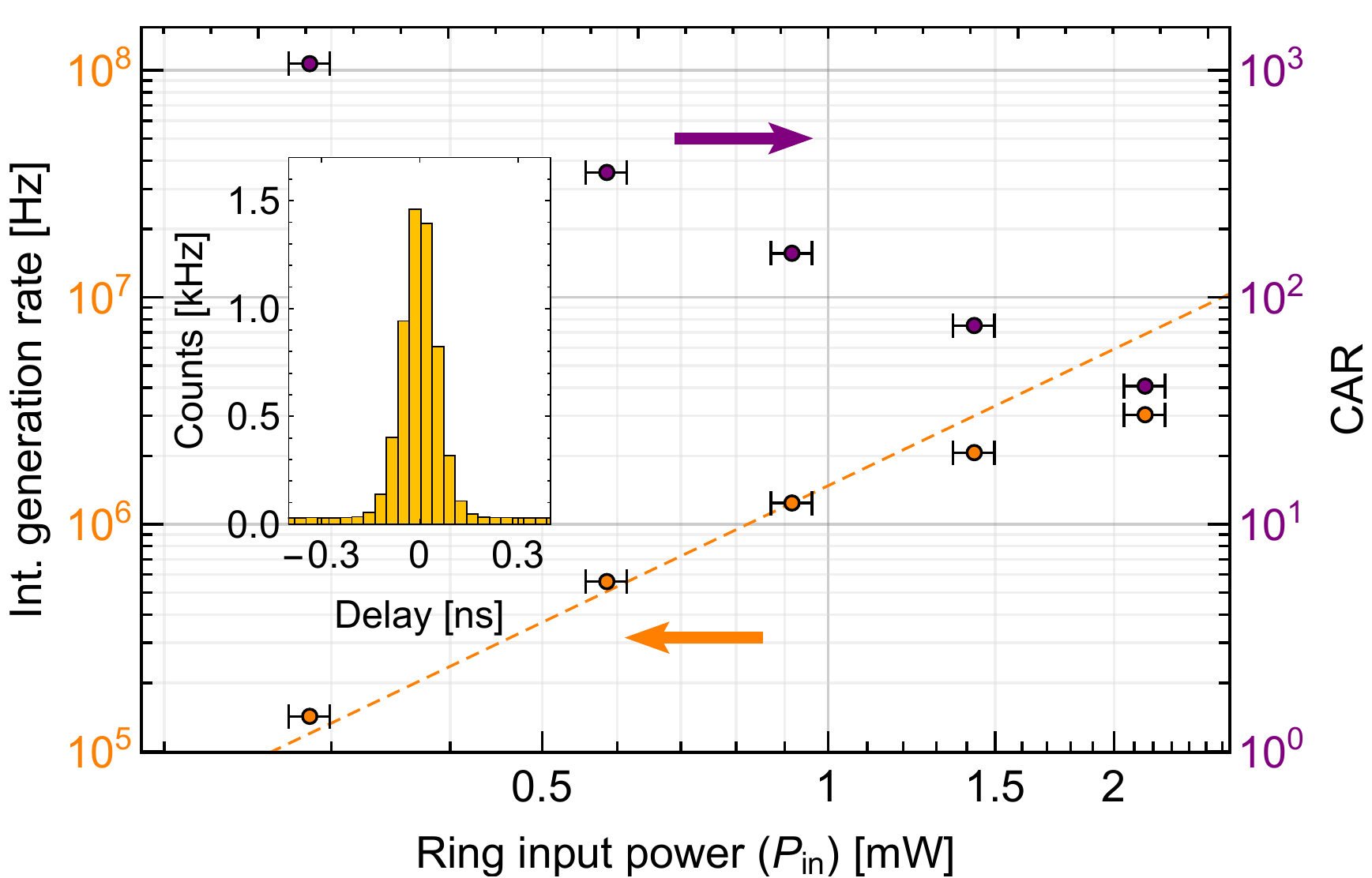}
    \caption{Internal signal-idler generation rate and CAR as a function of the cavity power at the ring input ($P_\text{in}$). The dashed orange line is a guide to the eye representing a perfect quadratic scaling. Horizontal error bars represent a $\pm 5\%$ uncertainty in the estimation of the cavity power; vertical error bars are smaller than the markers. Inset: representative coincidence histogram for $P_\text{in}$ = 2.157 mW; time bins are 35 ps wide.}
    \label{fig:coincidences}
\end{figure}

In Figure \ref{fig:coincidences} we report the result of coincidence measurements as a function of the pump power at the ring input.
The internal pair generation rate was inferred from the measurement of the propagation losses and the SSPDs efficiency for signal and idler, which are in total -13.1 dB and -14.2 dB, respectively.
The power scaling of the pair generation rate is quadratic for lower cavity powers, thus confirming that SFWM is the dominant nonlinear process inside the microring \cite{Helt2010}.
Remarkably, although some saturation is present, the rate keeps increasing also above 1 mW, when the ring resonance is shifted by more than one linewidth, without the need of any active feedback of the pump on the ring resonance \cite{savanier2015optimizing}.
In Figure \ref{fig:coincidences} we also report the coincidence to accidental ratio (CAR), defined as the ratio between the background-subtracted coincidences within the FWHM of the peak and the background coincidences over the same interval.

\textcolor{red}{As expected the CAR is strongly reduced at high generation rates; however, all the values are about two orders of magnitude lower than what would be predicted from the effect of the emission of multiple pairs alone \cite{Takesue2010}. This clearly indicates the presence of excess noise from other sources, such as Raman emission.}

The spectral correlations of the photons generated by parametric fluorescence in a micro ring resonator depend strongly on the spectral profile of the pump \cite{Helt2010,Christensen18}.
As shown in Figure \ref{fig:thermoshift}, in our case the pump spectrum is a function of the circulating power, and consequences on the correlations of the generated photons are expected.
While a complete characterization of the spectral properties of the state requires its tomography, one can have important information about spectral correlations by looking at the joint spectral density (JSD) $|\phi(\omega_1,\omega_2)|^2$, where $\phi(\omega_1,\omega_2)$ is the biphoton wavefunction \cite{Eckstein:LasPhotRev:2014}. 
We reconstructed the JSD by stepping the cw external tunable laser (Fig. \ref{fig:setup}) across the signal resonance and collecting the spectrum of the idler generated by stimulated FWM from the Through port of the ring.
The spectrum is measured with a resolution of 1 pm using a tunable FP filter, with its output sent to a SSPD.
Compared to energy-resolved coincidence experiments \cite{Avenhaus:OpLett:09}, this approach greatly decreases acquisition times by relying on the connection between stimulated and spontaneous parametric processes \cite{Liscidini:2013}.

\begin{figure*}[t]
    \centering
    \includegraphics[width=0.9\textwidth]{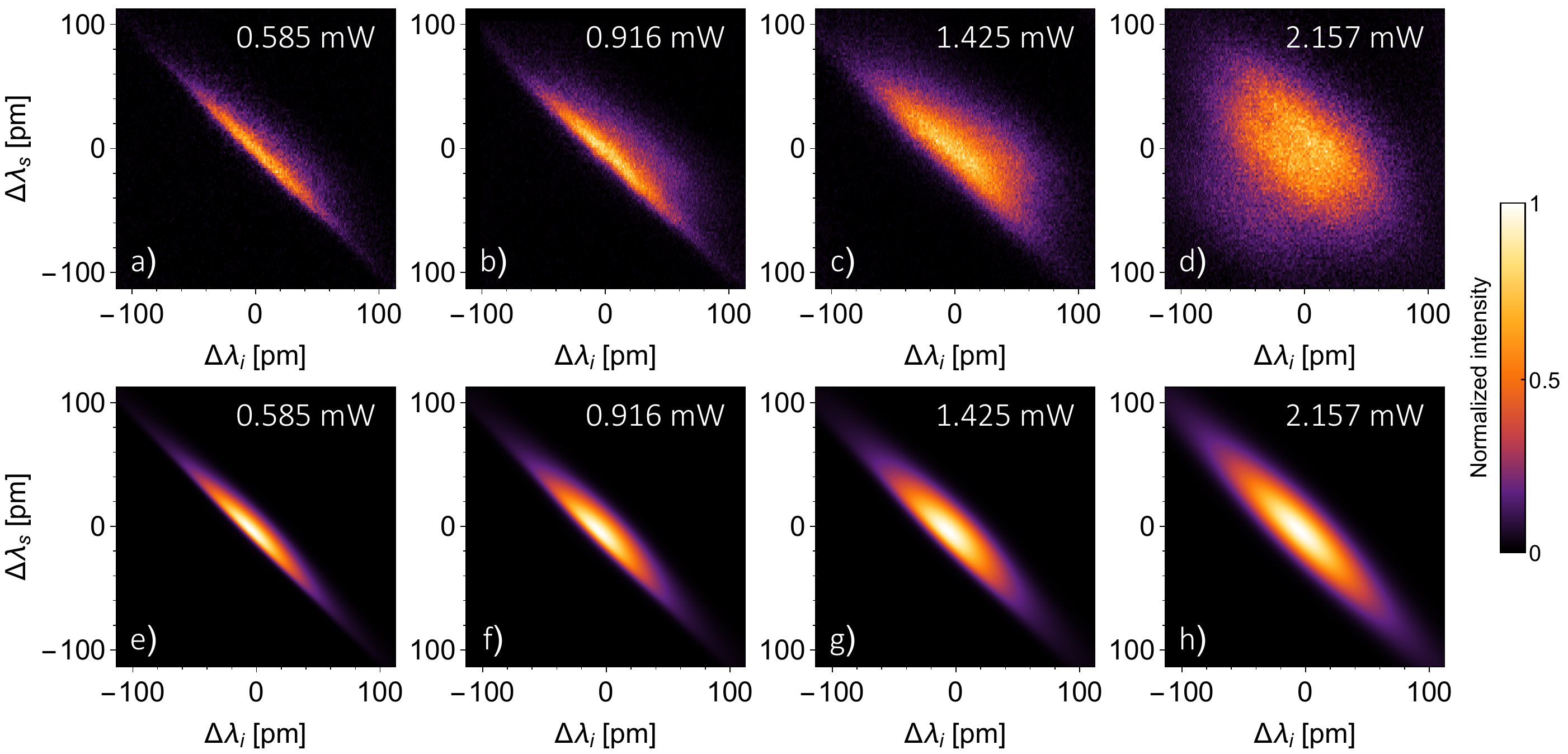}
    \caption{a-d) Joint spectral density (JSD) \textcolor{red}{reconstruction} for different cavity powers. The horizontal and vertical resolution are 1 and 2~pm respectively.
    e-h) Simulation of the JSD as expected from the measured pump lasing spectra, assuming coherent radiation and absence of nonlinear losses. The associated Schmidt numbers vary between 4.07 (e) to 2.75 (h).}
    \label{fig:JSD}
\end{figure*}


The results of the measurements are reported in the top row of Figure \ref{fig:JSD}, where we show JSDs for different values of the ring input power. As the pump power is increased, one observes a reduction of the spectral correlations of the generated photons. This is consistent with the expected pump spectral broadening shown in Fig. \ref{fig:thermoshift}. For pump powers larger than 1.4 mW (\ref{fig:JSD}-d)), the situation is more complex and the trend is not entirely clear. In the bottom row of Fig. \ref{fig:JSD}, we show the theoretical JSDs, which have been calculated by modelling SFWM in a microring resonator \textcolor{red}{taking into account the increased linewidth due to TPA at high power}, and considering a coherent pump with spectrum equal to the ones of Figure \ref{fig:thermoshift}. At low pump powers,  (Fig. \ref{fig:JSD} e)-g)), we find a good agreement with the experimental results, indicating that in this regime the systems is essentially behaving like an ideal one.
However, there is a substantial disagreement at the highest values of the pump power (Fig. \ref{fig:JSD} d) and h)).

\textcolor{red}{The degree of entanglement between photon pairs can be calculated by performing the Schmidt decomposition of the biphoton wavefunction, under the assumption that the biphoton state is pure \cite{law2000continuous}.
We find that the Schimidt number associated to the theoretical JSDs shown in Figure \ref{fig:JSD} decreases from a maximum of 4.07 to a minimum of 2.75 as the pump power is increased.}\\



In this work we have characterized the photon pair state generated by a silicon microring resonator in a self-pumped configuration, observing that the degree of correlation 
\textcolor{red}{is limited by the multimodal nature of the lasing radiation} and it is decreased with the amplifier current.
While a good agreement is present at lower powers between theoretical models and experimental data, the same is not true at higher powers, where, additionally, saturation of the pair generation rate occurs together with a broadening of the lasing radiation.
Further investigation is required to explain this divergence, here attributed to nonlinear losses induced through TPA and to gain saturation effects in the amplifier.
\textcolor{red}{In a similar scheme, an etalon has been placed along the self-pumping loop in order to obtain single-mode lasing operation and thus a much higher degree of entanglement \cite{garrisi2020electrically};
however, the etalon had to be tuned on the microring pump resonance, thus requiring an active feedback to maintain lasing at high power.
}
Here, instead, we demonstrated that the emission of pairs is maintained without active feedback on the pump even at regimes for which the laser pump resonance is shifted by more than one linewidth.\\

\noindent\textbf{Funding.}
This work was supported by the Italian Ministry of Education, University and Research (MIUR): “Dipartimenti di Eccellenza Program (2018–2022)”, Department of Physics, University of Pavia.\\

\bibliography{citations}

\end{document}